\documentclass[prd,aps,showpacs,tightenlines,nofootinbib,preprint]{revtex4}
\usepackage{amssymb}

\newcommand{\bear}{\begin{array}}  \newcommand{\eear}{\end{array}}
\newcommand{\bea}{\begin{eqnarray}}  \newcommand{\eea}{\end{eqnarray}}
\newcommand{\beq}{\begin{equation}}  \newcommand{\eeq}{\end{equation}}
\newcommand{\bef}{\begin{figure}}  \newcommand{\eef}{\end{figure}}
\newcommand{\bec}{\begin{center}}  \newcommand{\eec}{\end{center}}

\newcommand{\bib}{\bibitem}

\def\APJ#1#2#3{Astrophys. J. {\bf #1}, #2 (19#3)}

\def\ARAA#1#2#3{Annu. Rev. Astron. Astrophys. {\bf#1}, #2 (19#3)}

\def\IJMPDD#1#2#3{Int. J. Mod. Phys. D {\bf #1}, #2 (20#3)}

\def\JL#1#2#3{JETP. Lett. {\bf #1}, #2 (19#3)}

\def\JHEPP#1#2#3{J. High Energy Phys. {\bf #1}, #2 (20#3)}

\def\MPLA#1#2#3{Mod. Phys. Lett. A {\bf #1}, #2 (19#3)}

\def\NPB#1#2#3{Nucl. Phys. {\bf B#1}, #2 (19#3)}

\def\PLB#1#2#3{Phys. Lett. B {\bf #1}, #2 (19#3)}
\def\PLBB#1#2#3{Phys. Lett. B {\bf #1}, #2 (20#3)}
\def\PL#1#2#3{Phys. Lett. {\bf #1}, #2 (19#3)}

\def\PRD#1#2#3{Phys. Rev. D {\bf #1}, #2 (19#3)}
\def\PRDD#1#2#3{Phys. Rev. D {\bf #1}, #2 (20#3)}

\def\PRL#1#2#3{Phys. Rev. Lett. {\bf#1}, #2 (19#3)}

\def\PRT#1#2#3{Phys. Rep. {\bf#1}, #2 (19#3)}
\def\PRTT#1#2#3{Phys. Rep. {\bf#1}, #2 (20#3)}

\def\PTPP#1#2#3{Prog. Theor. Phys. {\bf #1}, #2 (20#3)}

\def\RPP#1#2#3{Rep. Prog. Phys. {\bf #1}, #2 (19#3)}

\def\RMPP#1#2#3{Rev. Mod. Phys. {\bf #1}, #2 (20#3)}

\def\Vec#1{\mbox{\boldmath $#1$}}
\def\Vecs#1{\mbox{\boldmath\tiny $#1$}}

\begin{document}

\title{Large-scale magnetic fields from dilaton inflation in noncommutative 
spacetime 
      }

\author{Kazuharu Bamba and J.\ Yokoyama}
\affiliation{Department of Earth and Space Science, Graduate School of
Science, Osaka University, Toyonaka 560-0043, Japan}

\date{\today}

\begin{abstract}
The generation of large-scale magnetic fields is studied in dilaton 
electromagnetism in noncommutative inflationary cosmology taking 
into account the effects of the spacetime uncertainty principle 
motivated by string theory.  
We show that it is possible to generate 
large-scale magnetic fields with sufficient strength 
to account for the observed fields in galaxies 
and clusters of galaxies through only adiabatic compression without 
dynamo amplification mechanism 
in models of power-law inflation based on spacetime noncommutativity 
without introducing a huge hierarchy between the dilaton's potential and 
its coupling to the electromagnetic fields.  
\end{abstract}

\pacs{98.80.Cq, 98.62.En}
\hspace{13.4cm} OU-TAP-224

\maketitle

\section{Introduction}
It is well established that magnetic fields with the field strength 
$\sim 10^{-6}$G and coherent scale $1-10$kpc exist in our galaxy and other 
galaxies (for detailed reviews see 
\cite{Kronberg1, Grasso, Widrow, Sofue, Giovannini0}).  
There is some evidence that they exit in galaxies at cosmological distances 
\cite{Kronberg2}.  Furthermore, in recent years magnetic fields in clusters of 
galaxies have been observed by means of the Faraday rotation measurements 
(RMs) of polarized electromagnetic radiation passing through an ionized 
medium \cite{Kim1}.  In general, the strength and the coherent scale are 
estimated as $10^{-7}-10^{-6}$G and 10kpc$-$1Mpc, respectively.  
It is very interesting and mysterious that magnetic fields in clusters of 
galaxies are as strong as galactic ones and that the coherence scale may be 
as large as $\sim$Mpc.  

Although galactic dynamo mechanisms \cite{EParker} have been proposed 
to amplify very weak seed magnetic fields up to $\sim 10^{-6}$G, it is only 
an amplification mechanism, and so requires initial seed magnetic fields to 
feed on.  
Furthermore, the effectiveness of the dynamo amplification mechanism 
in galaxies at high redshifts or clusters of galaxies 
is not well established.  
Hence the origin of the magnetic fields on large scales and/or at high 
redshift has been an open question.  

Proposed generation mechanisms of seed magnetic fields fall into two broad 
categories.  One is astrophysical processes, and 
the other is cosmological processes in the early Universe, 
\textit{e.g.}, the first-order cosmological electroweak phase 
transition (EWPT) \cite{Baym} or quark-hadron phase transition (QCDPT) 
\cite{Quashnock}.  
However, 
it is hardly possible for astrophysical processes to generate 
magnetic fields on megaparsec scales.  
Moreover, it is also difficult to make the mechanisms at the cosmological 
phase transitions operate on these scales today, which are much larger than 
the horizon scale at the epoch of the field generation 
(see also \cite{Durrer}).  

The most natural origin of such a large-scale magnetic field would be 
electromagnetic quantum fluctuations generated in the inflationary stage 
\cite{Turner} (for a comprehensive introduction to inflation see 
Refs. \cite{Linde1,Kolb}).  
This is because inflation naturally produces effects on very large scales, 
larger than Hubble horizon, starting from microphysical processes 
operating on a causally connected volume.  
However, there is a serious obstacle on the way of this nice scenario as argued below.  

It is well known that quantum fluctuations of massless scalar and tensor 
fields in the inflationary stage create 
considerable density inhomogeneities \cite{Guth} 
or relic gravitational waves \cite{Starobinsky,Rubakov}.  
This is because these fields are not conformally invariant even though 
they are massless.  Since the Friedmann-Robertson-Walker (FRW) metric 
usually considered is conformally flat, cosmic expansion does not 
induce particle production if the underlying theory is conformally invariant 
\cite{Parker}.  The classical electrodynamics is conformally invariant.  
Hence large-scale electromagnetic fluctuations could not be generated 
in cosmological background.  
If the origin of large-scale magnetic fields in 
clusters of galaxies is electromagnetic quantum fluctuations generated and 
stretched in the inflationary stage, the conformal invariance must have been 
broken at that time.  
Several breaking mechanisms therefore have been proposed 
\cite{Turner,Ratra,Scalar,Dolgov2,Bertolami}. 

Recently we have studied the following model of the dilaton electromagnetism 
\cite{Bamba}.  
In addition to the inflaton field $\phi$ 
we assumed the existence of the dilaton 
field $\Phi$ with a potential 
$
V[\Phi] = \bar{V} \exp(-\tilde{\lambda} \kappa \Phi), 
$ 
where $\bar{V}$ is a constant, and introduced the following coupling 
in the electromagnetic part of the model Lagrangian, 
\begin{eqnarray}
{\mathcal{L}}_{\mathrm{EM}}  &=& 
                -\frac{1}{4} f(\Phi) F_{\mu\nu}F^{\mu\nu}, 
\label{eq:1} \\[3mm]
f(\Phi) &=& \exp(\lambda \kappa \Phi), 
\label{eq:2}
\end{eqnarray} 
where 
$F_{\mu\nu}={\partial}_{\mu}A_{\nu} - {\partial}_{\nu}A_{\mu}$ 
is the electromagnetic field-strength tensor, 
$\lambda$ and $\tilde{\lambda}$ are dimensionless constants, 
and ${\kappa}^2 \equiv 8\pi/{M_{\mathrm{Pl}}}^2$ with 
$M_{\mathrm{Pl}} = G^{-1/2} = 1.2 \times 10^{19}$GeV being the Planck mass.  
We use units in which $k_\mathrm{B} = c = \hbar = 1$ and adopt 
Heaviside-Lorentz units of electromagnetism.  
Such coupling is reasonable in the light of indications in higher-dimensional 
theories including string theory.  
The coupling was first analyzed by Ratra \cite{Ratra}.  
In his model, however, the inflaton and the dilaton were identified and 
only the case the dilaton freezes at the end of inflation was considered.  
We therefore considered a more realistic situation that 
the dilaton continues its evolution along with the exponential potential 
even after reheating but is finally stabilized when it feels other 
contributions to its potential, say, from gaugino condensation \cite{GC}
that generates a potential minimum \cite{Barreiro, Seto}.  
As it reaches there, the dilaton starts oscillation with mass $m$ and 
finally decays into radiation with or without significant entropy production.  
As a result we have shown magnetic fields with the current strength 
as large as $10^{-10}$G on cluster scale or even larger scale 
could be generated, but for this 
to be the case we had to introduce a huge hierarchy between 
the coupling constant of the dilaton to the electromagnetic field $\lambda$ 
and the coupling one $\tilde\lambda$ of the dilaton potential, 
$\lambda/\tilde{\lambda} \approx 400$.  

Note that the existence of the dilaton under discussion is motivated by 
higher dimensional theories including string theory. The purpose of the 
present paper is to argue that if we take another prediction of string 
theory, namely spacetime uncertainty relation, we can solve the above 
huge hierarchy problem as well.  
As emphasized by Yoneya \cite{Yoneya}, 
the stringy spacetime uncertainty relation (SSUR) is {\it not} a modification 
of the ordinary energy-time uncertainty relation in the framework of quantum 
mechanics, 
but simply a {\it reinterpretation} in terms of strings.  
Hence the SSUR is likely to be very universal in string theories.  
It is therefore natural and important to take into account 
both of the two consequences of string theory, the dilaton and the SSUR, 
simultaneously to the problem of the generation of primordial magnetic fields 
in the high energy regime of the early Universe.  

The rest of the paper is organized as follows.  
In Sec.\ II we review our previous model and outline the result and 
the hierarchy problem to be solved.  
In Sec.\ III we consider the effects of the SSUR on the power spectrum of 
magnetic fields and apply the noncommutative effects on fluctuations to 
our previous model to solve the above huge hierarchy problem.  
Finally, Sec.\ IV is devoted to conclusion.

\section{Generation of magnetic fields in dilaton electromagnetism}

\subsection{U(1) gauge field in exponential inflation}
From (1) 
the equation of motion for 
the U(1) gauge field $A_{\mu}$ in the Coulomb gauge, 
$A_0(t,\Vec{x}) = 0$ and ${\partial}_jA^j(t,\Vec{x}) =0$, reads 
\begin{eqnarray}
  \ddot{A_i}(t,\Vec{x})
 + \left( H + \frac{\dot{f}}{f} \right)
               \dot{A_i}(t,\Vec{x})
- \frac{1}{a^2}{\partial}_j{\partial}_jA_i(t,\Vec{x}) = 0, 
\label{eq:3} 
\end{eqnarray}   
in the spatially flat Robertson-Walker Universe 
$ds^2 = -dt^2 + a^2(t) d{\Vec{x}}^2$, 
where $H$ is the Hubble parameter and $a$ is the cosmic scale factor.  
Through the canonical quantization the expression for $A_i(t,\Vec{x})$ 
is given by 
\begin{eqnarray} 
  A_i(t,\Vec{x}) = \int \frac{d^3 k}{{(2\pi)}^{3/2}}
  \left[ \hspace{0.5mm} \hat{b}(\Vec{k}) 
        A_i(t,\Vec{k})e^{i \Vecs{k} \cdot \Vecs{x} }
       + {\hat{b}}^{\dagger}(\Vec{k})
       {A_i}^*(t,\Vec{k})e^{-i \Vecs{k} \cdot \Vecs{x}} \hspace{0.5mm} \right],
\label{eq:4} 
\end{eqnarray}
where $\hat{b}(\Vec{k})$ and ${\hat{b}}^{\dagger}(\Vec{k})$ 
are the annihilation and creation operators which satisfy 
\begin{eqnarray} 
\left[ \hspace{0.5mm} \hat{b}(\Vec{k}), {\hat{b}}^{\dagger}({\Vec{k}}^{\prime}) \hspace{0.5mm} \right] = 
{\delta}^3 (\Vec{k}-{\Vec{k}}^{\prime}), \hspace{5mm}
\left[ \hspace{0.5mm} \hat{b}(\Vec{k}), \hat{b}({\Vec{k}}^{\prime})
\hspace{0.5mm} \right] = 
\left[ \hspace{0.5mm} 
{\hat{b}}^{\dagger}(\Vec{k}), {\hat{b}}^{\dagger}({\Vec{k}}^{\prime})
\hspace{0.5mm} \right] = 0.
\label{eq:5} 
\end{eqnarray}
Here $\Vec{k}$ is comoving wave number, and $k$ denotes its amplitude 
$|\Vec{k}|$.  
It follows from Eq.\ (\ref{eq:3}) that the Fourier modes 
$A_i(t,k)$ satisfy the equation, 
\begin{eqnarray} 
\ddot{A_i}(t,k) + \left( H + \frac{\dot{f}}{f} \right)
               \dot{A_i}(t,k) + \frac{k^2}{a^2} A_{i}(t,k) = 0,  
\label{eq:6} 
\end{eqnarray} 
and that the normalization condition for $A_i (t,k)$ reads
\begin{eqnarray} 
A_i(t,k){\dot{A}}_j^{*}(t,k) - {\dot{A}}_j(t,k){A_i}^{*}(t,k)
= \frac{i}{f(\Phi)a(t)} \left( {\delta}_{ij} - \frac{k_i k_j}{k^2 } \right).
\label{eq:7} 
\end{eqnarray}
For convenience in finding the solutions of Eq.\ (\ref{eq:6}), 
we introduce the following approximate form as the expression of $f$.  
\begin{eqnarray}
f(\Phi) = f[\Phi(t)] = f[ \hspace{0.5mm} \Phi (a(t)) \hspace{0.5mm} ]
        \equiv \bar{f}a^{\beta-1}, 
\label{eq:8}
\end{eqnarray} 
where $\bar{f}$ is a constant and $\beta$ is a parameter.  

In slow-roll exponential inflation models with 
$a(t) \propto {e}^{ H_{\mathrm{inf}} t }$, 
the model parameter $\beta$ is given by 
\begin{eqnarray}
&& \beta \approx  1 + {\tilde{\lambda}}^2 w X, 
\hspace{5mm}
 X \equiv \frac{\lambda}{ \tilde{\lambda} }, 
\label{eq:9} \\[3mm]
&& w     \equiv   \frac{ V[\Phi]}{{\rho}_{\phi} },
\label{eq:10}
\end{eqnarray} 
where ${\rho}_{\phi} \cong \mathrm{const}$ 
is the energy density of the inflaton $\phi$ 
and $w \ll 1$ because we have assumed that 
during slow-roll inflation the cosmic energy density is dominated 
by the inflaton potential 
and the energy density of the dilaton is negligible.  
Even if the dilaton was rapidly evolving at the onset, its kinetic energy 
would soon be dissipated, and it is frozen to a value satisfying 
$V^{\prime\prime} [\Phi] \lesssim {H_\mathrm{inf}}^2$.  
Thus $\beta$ takes a practically constant value.  
Consequently, the solution of Eq.\ (\ref{eq:6})  
satisfying Eq.\ (\ref{eq:7}) with $H = H_{\mathrm{inf}}$ 
is given by 
\begin{eqnarray}
A_i(k,a) = \sqrt{\frac{\pi}{4H_{\mathrm{inf}}af(a)}} 
     H_{\beta/2}^{(1)} \left( \frac{k}{a H_{\mathrm{inf}}} \right) 
     e^{i(\beta+1)\pi/4},   
\label{eq:11}
\end{eqnarray}
where we have determined the constants of integration in the general solution 
of Eq.\ (\ref{eq:6}) by requiring that the vacuum 
reduces to the one in Minkowski spacetime at the short-wavelength limit.  

The energy density of the large-scale magnetic fields 
can evaluated using (\ref{eq:11}) if we specify cosmic evolution after 
inflation.  Here we adopt the following scenario.  
After inflation, the inflaton potential is instantaneously converted into 
radiation and then the Universe is reheated immediately at $t=t_\mathrm{R}$.  
On the other hand, 
even after reheating the dilaton continues its evolution along with 
the exponential potential $V[\Phi]$ for a while, but is finally stabilized 
when it feels other contributions to its potential, say, from gaugino 
condensation \cite{GC} that generates a potential minimum 
\cite{Barreiro, Seto}.  
As it reaches there, the dilaton starts oscillation with mass $m$ and 
finally decays into radiation with entropy production, and hence the 
energy density of magnetic fields is diluted.  
Then the energy density of the magnetic field on a comoving scale 
$L=2\pi/k$ 
at the present time 
is given by 
\begin{eqnarray}
{\rho}_B(L,t_0) 
   &=&
  \frac{2^{|\beta|-3}}{{\pi}^3} {\Gamma}^2 \left( \frac{|\beta|}{2} \right)
   {H_{\mathrm{inf}}}^4 
   \left( \frac{a_{\mathrm{R}}}{a_0}  \right)^4 
  { \left( \frac{k}{a_{\mathrm{R}} H_{\mathrm{inf}}} \right) }^{-|\beta|+5} 
\nonumber  \\[2.5mm]
&& \hspace{50mm} \times 
  \exp \left( \hspace{0.5mm}  - \tilde{\lambda} \kappa {\Phi}_\mathrm{R}
          X  \hspace{0.5mm} \right) 
  \left( \Delta S  \right)^{-4/3}, 
\label{eq:12} \\[5mm]
\Delta S
&\approx&
 \left( \frac{\bar{V}}{{\rho}_{\phi}}  \right)
 \left( \frac{2H_{\mathrm{inf}}}{m} \right)^2
 \left( \frac{M_\mathrm{Pl}}{m} \right), 
\label{eq:13} 
\end{eqnarray} 
where 
${\Phi}_\mathrm{R}$ is the dilaton field amplitude at the end of inflation and 
$\Delta S$ is the entropy ratio after and before dilaton decay.  
Here the suffixes `R' and `0' represent the quantities at the end of inflation 
$t_\mathrm{R}$ 
and the present time $t_0$, respectively.  
From Eq.\ (\ref{eq:12}) we see that the large-scale magnetic fields have a scale-invariant spectrum when $|\beta| = 5$ \cite{Bamba}.  

After a number of consistency arguments 
we have found that the magnetic field could be 
as large as $10^{-10}$G even with the entropy increase 
factor $\Delta S \sim 10^6$, which is the ratio of the entropy per 
comoving volume after the dilaton decay to that before decay, 
provided that the energy scale of inflation is maximal and 
the spectrum of resultant magnetic field is close to the 
scale-invariant or the red one, namely, $\beta \gtrsim 5$ \cite{Bamba}.  
It follows from Eq.\ (\ref{eq:12}) that the magnetic field strength on 1Mpc 
scale at the present time is 
\begin{eqnarray}
&& \hspace{-12mm} B(1\mathrm{Mpc},t_0) \approx 
3.5 \times 10^{-12} \times 2^{X/200} 
\Gamma \left( \frac{X}{200} + \frac{1}{2} \right) 
\left( \frac{H_{\mathrm{inf}}}{H_{\mathrm{max}}} \right) 
\nonumber  \\[2.5mm]
&& \hspace{1.5cm}
\times
\exp \left[ \left( 53.5 + \frac{1}{2} 
\ln \left( \frac{H_{\mathrm{inf}}}{H_{\mathrm{max}}} \right) 
\right) 
            \left( \frac{X}{200} - 2 \right)  
           - \frac{1}{2} \tilde{\lambda} \kappa {\Phi}_\mathrm{R} X
\right] 
\left( \Delta S  \right)^{-2/3} \hspace{1mm} \mathrm{G}, 
\label{eq:14} 
\end{eqnarray} 
where we have taken 
$w =0.01$, ${\tilde{\lambda}} \sim \mathcal{O}(1)$, and then 
$\beta \approx  1 + X/100$.  
Here $H_{\mathrm{max}} \equiv 2.4 \times 10^{14} \mathrm{GeV}$ 
is the maximum possible value of $H_{\mathrm{inf}}$ imposed by 
the amplitude of the tensor perturbations \cite{Rubakov,Abbott}.  
If the generated magnetic fields are as large as $10^{-10}$G, 
the observed fields in galaxies and clusters of galaxies could be explained 
through only adiabatic compression without dynamo amplification mechanism.  
Incidentally, Caprini, Durrer, and Kahniashvili \cite{Caprini} have recently 
investigated the effect of gravitational waves induced by a possible 
helicity-component of a primordial magnetic field on CMB temperature 
anisotropies and polarization.  
According to them, the effect could be sufficiently large to be observable 
if the spectrum of the primordial magnetic field is close to 
scale invariant and if its helical component is stronger than 
$\sim 10^{-10}$G.  
Hence our scenario may be observationally testable 
(see also \cite{Lewis}). 

On the other hand, 
the seed field required for the dynamo mechanism, $B \sim 10^{-22}$G, 
could be accounted for 
even when $\Delta S$ is as large as $10^{24}$ if model parameters are 
chosen appropriately to realize nearly scale-invariant spectrum.  

The serious problem is that the model parameters should be so
chosen that the spectrum of generated magnetic field should not be too
blue but close to the scale-invariant or the red one, which is realized
only if a huge hierarchy exists between $\lambda$ and $\tilde\lambda$, 
namely, 
$X = \lambda/\tilde{\lambda}$ should be extremely larger than unity.  
For example, 
if we take $w =0.01$ and ${\tilde{\lambda}} \sim \mathcal{O}(1)$, 
we must have $X \equiv \lambda/\tilde{\lambda}$ as large as $ \approx 400$ 
so that 
the amplitude of the generated large-scale magnetic field could be 
sufficiently large.  
This may make it difficult to motivate this type of model in realistic high 
energy theories.

\subsection{The case of power-law inflation}
If we adopt power-law inflation models instead of exponential inflation 
with the following exponential inflaton potential,  
\begin{eqnarray}
  U[\phi] = \bar{U} \exp(-\zeta \kappa \phi), 
\label{eq:15}
\end{eqnarray}
we can relax the constraint on $X$ to a limited extent.  
Here $\bar{U}$ is a constant and $\zeta$ is a dimensionless constant, 
the spectral index of curvature perturbation $n_s$ 
is given by 
\begin{eqnarray}
   n_s - 1 = -6 {\epsilon}_U + 2 {\eta}_U = - {\zeta}^2, 
\label{eq:16} 
\end{eqnarray}
with 
\begin{eqnarray}
  {\epsilon}_U \equiv \frac{1}{2{\kappa}^2}  
      \left( \frac{U^{\prime}}{U} \right)^2, 
\hspace{5mm} 
  {\eta}_U \equiv \frac{1}{{\kappa}^2}  
      \left( \frac{U^{\prime\prime}}{U} \right), 
\label{eq:17} 
\end{eqnarray}
where primes denote derivatives with respect to the inflaton field $\phi$.  
According to the first year Wilkinson Microwave Anisotropy Probe 
(WMAP) data \cite{Peiris}, $n_s \geq 0.93$, and hence 
$\zeta \leq 0.26$.  In this case the scale factor in the inflationary stage 
is given by $a(t) \propto t^p$, where $p=2/{\zeta}^2 \geq 29$.  

In this background, 
if power-law inflation lasts for a sufficiently long time, the dilaton 
will settle to the scaling solution \cite{Barreiro} where 
$U^{\prime\prime}[\phi] \approx {H}^2$ 
with $H = p/t$.  
Hence the solution of the dilaton in this regime is given by 
\begin{eqnarray}
   \Phi = \frac{2}{\tilde{\lambda}\kappa}
  \ln \left( \hspace{0.5mm}  
     \frac{ \sqrt{\bar{V}} \tilde{\lambda}\kappa t }{p}
      \right).  
\label{eq:18}
\end{eqnarray}  
Then we find $\beta$ is constant given by 
\begin{eqnarray}
\beta = \frac{2X}{p} + 1. 
\label{eq:19}
\end{eqnarray} 
In this case, the solution of Eq.\ (\ref{eq:6}) 
is given by 
\begin{eqnarray}
A_i(k,a) &=& \sqrt{ \frac{p \pi}{4(p-1)Haf(a)} }  
    H_{\tilde{\beta}/2}^{(1)} 
    \left( \frac{pk}{(p-1)aH}  \right) 
    e^{i(\tilde{\beta}+1)\pi/4}, 
\label{eq:20} \\[3mm]
\tilde{\beta} &\equiv& 1 + \frac{p}{p-1} \left( \beta - 1 \right) 
= 1+\frac{2X}{p-1}.  
\label{eq:21}
\end{eqnarray}
The energy density of the large-scale magnetic fields 
on a comoving scale $L=2\pi/k$ at the present time 
in the above power-law inflation models, 
$\tilde{ {\rho}_B }(L,t_0)$, 
is given by 
\begin{eqnarray}
\tilde{ {\rho}_B }(L,t_0) 
   &=&
  \frac{2^{|\tilde{\beta}|-3}}{{\pi}^3} 
  {\Gamma}^2 \left( \frac{|\tilde{\beta}|}{2} \right)
  \left( \frac{p-1}{p} \right)^{ |\tilde{\beta}| - 1}
   {H_{\mathrm{R}}}^4 
   \left( \frac{a_{\mathrm{R}}}{a_0}  \right)^4 
  { \left( \frac{k}{a_{\mathrm{R}} H_{\mathrm{R}}} \right) }^
                                                     {- |\tilde{\beta}| + 5} 
\nonumber  \\[2.5mm]
&& \hspace{60mm} \times 
  \exp \left( \hspace{0.5mm}  - \tilde{\lambda} \kappa {\Phi}_\mathrm{R}
          X  \hspace{0.5mm} \right) 
  \left(  \Delta \tilde{S}   \right)^{-4/3}, 
\label{eq:22} \\[5mm]
\Delta \tilde{S} 
&\approx&
 \left( \frac{\bar{V}}{{\rho}_{\phi}}  \right)
 \left( \frac{2H_{\mathrm{R}}}{m} \right)^2
 \left( \frac{M_\mathrm{Pl}}{m} \right), 
\label{eq:23} 
\end{eqnarray} 
where $H_{\mathrm{R}}$ is the Hubble parameter at the end of inflation.  
Since $p \gg 1$ as noted above, we find from 
Eqs.\ (\ref{eq:12}), (\ref{eq:13}), and (\ref{eq:21})$-$(\ref{eq:23}) 
that $\tilde{\beta} \approx \beta$, 
which means $\tilde{ {\rho}_B }(L,t_0) \approx {\rho}_B(L,t_0)$ 
by identifying $H_{\mathrm{R}}$ with $H_{\mathrm{inf}}$. 
Therefore, if $\beta \approx 5$, we find $X \approx 2p \geq 58$ \cite{Bamba}.  
Consequently,  
although some progress has been made to lower the required value of $X$ by 
adopting power-law inflation, it is far from sufficient because $X$ should 
still be much larger than unity in order that 
the amplitude of the generated magnetic fields could be sufficiently large 
at the present time.  
This is because power-law inflation models are hardly distinguishable 
from exponential inflation under the constraint imposed by WMAP data 
as far as the evolution of the dilaton is concerned.  
Because we cannot expect in realistic high energy theories that 
a huge hierarchy exists between $\lambda$ and $\tilde\lambda$, 
this is a serious problem of our previous model to be solved.

\section{Generation of magnetic fields 
with stringy spacetime uncertainty relation}
In this section, we consider a possible solution to the above 
huge hierarchy between $\lambda$ and $\tilde\lambda$ in our model. 
In recent years 
the effect of the stringy spacetime uncertainty relation (SSUR) 
\cite{Yoneya} 
\begin{eqnarray}
\Delta t \Delta x_{\rm phys} \geq L_s^2,
\label{eq:24} 
\end{eqnarray}
on metric perturbations in the early Universe have been investigated 
\cite{Brandenberger, Tsujikawa, Huang}, 
where $t$ and $x_{\rm phys}$ are the physical spacetime coordinates and
$L_s$ is the string scale.  
In the presence of the cosmic expansion, long-wavelength perturbations 
observable today emerged from the string-region in the early Universe, 
hence string-scale physics might leave an imprint 
on the primordial spectrum of metric perturbations.

\subsection{Application of noncommutative effects on density 
fluctuations to power-law inflation model}

The SSUR is compatible with a homogeneous background, 
but it leads to changes in the action for the metric fluctuations.  
Both scalar and tensor metric fluctuations can be described by the action 
of a free scalar field $\varphi$ on the classical expanding background.  
Brandenberger and Ho have first studied 
the modified action for the cosmological perturbations in noncommutative 
spacetime, where they have assumed that matter is dominated by a single 
scalar field \cite{Brandenberger}.  
We assume the spatially flat Friedmann-Robertson-Walker (FRW) spacetime 
and introduce a time coordinate $\tau$ so that the metric is 
\begin{eqnarray}
ds^2 = -dt^2 + a^2(t) d{\Vec{x}}^2
= -a^{-2}(\tau)d\tau^2 + a^2(\tau) d{\Vec{x}}^2.
\label{eq:25} 
\end{eqnarray} 
In this case, in terms of the Fourier transform of $\varphi$ 
\begin{eqnarray}
\varphi(\tau, \Vec{x}) = V^{1/2} \int_{k < k_{\mathrm{max}}(\tau)} 
\frac{d^3k}{{(2\pi)}^{3/2}}
\varphi_k(\tau) e^{i \Vecs{k} \cdot \Vecs{x}},
\label{eq:26} 
\end{eqnarray}
where $V$ is the total spatial coordinate volume, and $k_{\mathrm{max}}$ is 
an upper bound on the comoving wave number 
\begin{eqnarray}
&&k \leq {k}_{\mathrm{max}} \equiv \frac{ {a}_{\mathrm{eff}} (\tau) }{L_s}, 
\label{eq:27} \\[3mm]
&&{{a}_{\mathrm{eff}}}^2 (\tau) \equiv 
        \left( \frac{\beta^+_k (\tau)}{\beta^-_k (\tau)}
\right)^{1/2} 
= a(\tau - L_s^2 k) a(\tau + L_s^2 k), 
\label{eq:28} \\[3mm]
&&\beta^{\pm}_k(\tau) = \frac{1}{2} 
\left[
a^{\pm 2}(\tau - L_s^2 k)+a^{\pm 2}(\tau + L_s^2 k)
\right],
\label{eq:29}
\end{eqnarray} 
the action led by the SSUR is 
\begin{eqnarray}
S_{\mathrm{SSUR}} = V \int_{k < k_{\mathrm{max}}}
d{\tilde\eta} d^3k \frac{1}{2} {z_k}^{2}(\tilde\eta) 
\left( {\varphi}^{\prime}_{-k} {\varphi}^{\prime}_{k}
       - k^2 {\varphi}_{-k} {\varphi}_{k}
\right),
\label{eq:30} 
\end{eqnarray}
where a prime denotes derivatives with respect to 
the new time coordinate $\tilde\eta$ defined by 
\begin{eqnarray}
\frac{d \tilde\eta}{d \tau} \equiv 
\left(\frac{\beta^-_k}{\beta^+_k}\right)^{1/2}
= a^{-2}_{\mathrm{eff}},
\label{eq:31} 
\end{eqnarray}
and
\begin{eqnarray}
&&{z_k}^2(\tilde\eta) = z^2 (\beta^-_k \beta^+_k)^{1/2}.   
\label{eq:32}
\end{eqnarray}
\if
Here $z$ is equivalent to $a\dot{{\phi}_\mathrm{b}}/H$ in the case of 
scalar fluctuations, where ${\phi}_\mathrm{b}$ denotes the background 
value of the scalar matter field, and $a$ in that of tensor ones, 
respectively.  
\fi

Spacetime noncommutativity at high energies in the early Universe leads to 
the following two effects.  
The first is a coupling between the fluctuation mode and the background which 
is nonlocal in time.  
The second is the appearance of a critical time for each mode at which the 
SSUR is saturated, and which is taken to be the time when the mode is 
generated in the vacuum state in the absence of cosmological expansion.  
The reason is as follows.  
The SSUR must be satisfied in order that a fluctuation mode with the comoving 
wavenumber $k$ should exist.  
As described in Eq.\ (\ref{eq:27}), 
an upper bound is therefore imposed on the comoving wavenumber 
$k \leq {k}_{\mathrm{max}} \equiv {a}_{\mathrm{eff}} / L_s $.  
Hence the SSUR is saturated for a particular comoving wavelength 
when the corresponding physical wavelength is equal to the string length.  
Consequently, 
fluctuation modes must be considered to emerge at this time.  
Here we assume that the amplitude of fluctuation modes at the time of 
generation is the same as that in the vacuum state 
in Minkowski spacetime.  
In a background spacetime with power-law inflation, 
these two effects lead to a suppression of power for large-wavelength 
modes, compared to the predictions of power-law inflation in 
standard general relativity.  
This is because these modes undergo a shorter period of squeezing than 
they do in the standard calculations in the commutative geometry, that is, 
large-scale modes, which correspond to higher energies earlier in inflation, 
are generated outside the Hubble radius owing to stringy effects, 
and hence experience less growth than the small-scale modes, which are 
generated inside the Hubble radius at lower energies, and evolve as in the 
standard case.  
There is a critical wavenumber $k_{{\rm crit}}$ such that for $k <
k_{\rm crit}$ the mode is generated on super-Hubble scales, and hence 
undergoes less squeezing during the subsequent evolution than it
does in commutative spacetime, on the other hand, for $k > k_{\rm crit}$ 
the mode is generated on scales inside the Hubble radius, and 
since the evolution of the mode after that is not different from that 
in the case of commutative spacetime, it follows immediately that 
the spectrum for $k \gg k_{\rm crit}$ is the same as 
that in the classical case.  
Consequently, the spectrum is blue-tilted for $k \ll k_{\rm crit}$
rather than red-tilted as it is in the power-law inflation
scenario in commutative spacetime.  
If the scale factor in the inflationary stage is given by 
\begin{eqnarray}
a(t) = \bar{a} t^p = \bar{\alpha} {\tau}^{p/(p+1)}, \hspace{5mm}  
\bar{\alpha} = \left[ (p+1)^p \bar{a} \right]^{1/(p+1)}, 
\label{eq:33}
\end{eqnarray}
where $\bar{a}$ and $\bar{\alpha}$ are constants, 
the critical wavenumber is given by \cite{Brandenberger} 
\begin{eqnarray}
k_{{\rm crit}} = {\bar{\alpha}}^{p+1} {L_s}^{p-1}.  
\label{eq:34}
\end{eqnarray}  
From now on we call the modes $k \gg k_{\rm crit}$ the UV ones and 
the modes $k \ll k_{\rm crit}$ the IR ones, respectively.  

The spectrum of cosmic microwave background (CMB) anisotropies 
predicted by the model in \cite{Brandenberger} has recently been calculated, 
and thus the prediction of loss of power for infrared modes has 
been quantified \cite{Tsujikawa, Huang}.  
In addition, Tsujikawa et al.\ \cite{Tsujikawa} have performed 
a likelihood analysis at various angular scales 
to find the best-fit values to the WMAP data of 
the cosmological parameters, including the power-law 
exponent $p$ which gives the time dependence on the scale factor, 
and the critical wavenumber when stringy effects become important.  
As a result they have shown that 
high energy stringy effects could account for some loss of power 
on the largest scales that may be indicated by recent WMAP data.  
Moreover, 
the best-fit value for the power-law exponent $p$ has been found 
to be $p \approx 14$, which is 
consistent with the result of \cite{Huang}, in which 
the likelihood value of $p$ is derived by using recent WMAP data 
on only two scales $k=0.05\,{\rm Mpc}^{-1}$ and $k=0.002\,{\rm Mpc^{-1}}$.  
By using the best-fit values, 
the string energy scale ${L_s}^{-1}$ 
has been estimated as ${L_s}^{-1} \approx {10}^{14}$GeV.  
Furthermore, according to their calculation, 
even a power exponent as small as $p \approx 5$ is consistent 
within the current errors.  

If we apply the above consequences of the SSUR in a background spacetime 
with power-law inflation to our previous model \cite{Bamba}, 
from  Eq.\ (\ref{eq:21}) and $p \approx 5$ 
it is expected that $X$ could be much smaller than in the case of 
power-law inflation in commutative geometry which requires 
$p \geq 29$.  
\if
If we apply the above consequences of the SSUR in a background spacetime 
with power-law inflation to our previous model \cite{Bamba}, 
from  Eq.\ (\ref{eq:19}) and $p \approx 5$ 
we find $X \approx \frac{5}{2} (\beta - 1)$, namely, 
$\lambda$ and $\tilde\lambda$ could be of the same order of magnitude.  
For example, from  Eqs.\ (\ref{eq:21}) and (\ref{eq:22}) 
we see that the large-scale magnetic fields have a scale-invariant spectrum 
when $\tilde{\beta} = 5$, namely, $\beta = 1 + 4(p-1)/p$.  
In this case, if $p \approx 5$, 
$X = \lambda / \tilde\lambda =2(p-1) \approx 8$.  
As a result we find that 
the resultant magnetic fields can have a nearly scale-invariant spectrum 
even in the case 
$\lambda$ and $\tilde\lambda$ are of the same order of magnitude, that is, 
$X = \lambda / \tilde\lambda \approx 8$, 
so that the amplitude of the generated magnetic fields could 
be as large as 
$10^{-10}$G on 1Mpc scale at the present time.  
Moreover, 
we find similarly that 
a sufficient magnitude of magnetic fields on 1Mpc scale at the present time 
for the galactic dynamo scenario, $B \sim 10^{-22}$G, could be generated 
even in the case $X = \lambda / \tilde\lambda \approx 6$.  
\fi

One may wonder if 
electromagnetic quantum fluctuations generated in the inflationary stage 
are also influenced by spacetime noncommutativity, 
so that the power for the long-wavelength modes should be suppressed.  
However, the megaparsec scale, in which we are particularly interested, 
is smaller than the above critical scale $2\pi/k_{{\rm crit}}$.  
In fact, according to Tsujikawa et al.\ \cite{Tsujikawa}, 
the best-fit value for the crossover scale of the power spectrum 
of density fluctuation, $k_{*}$, which satisfies 
$k_{*} \gg k_{\mathrm{crit}}$, 
has been found to be 
$2\pi/k_{*} \approx 2.7 \times 10^2$Mpc.  
\if
In fact, according to Tsujikawa et al.\ \cite{Tsujikawa}, 
the best-fit value for a scale $k_{*} ( \gg k_{\mathrm{crit}})$ 
has been found to be 
$2\pi/k_{*} \approx 2.7 \times 10^2$Mpc.  
\fi
Hence the megaparsec scale 
fluctuations at the present time is the UV modes.  
It is therefore expected that 
the SSUR has no significant effect on the megaparsec scale fluctuations 
at the present time 
and that the amplitude of the generated magnetic field on 1Mpc scale is 
the same as that in the case of commutative spacetime.  
In next subsection, to confirm the above expectation 
we consider the modified power spectrum of magnetic fields 
and the strength of the generated magnetic field on 1Mpc scale 
at the present time in noncommutative spacetime.

\subsection{
Noncommutative modifications to the power spectrum of magnetic fields
}
As noted in the previous subsection, 
not only scalar and tensor metric fluctuations but also 
electromagnetic quantum fluctuations generated in the inflationary stage 
are also influenced by spacetime noncommutativity, 
so that the power for the large-wavelength modes should be suppressed.  
We therefore consider the modified power spectrum of magnetic fields.  

To begin with, 
we consider the action for the Fourier modes $A_i (k, \eta)$ 
of the U(1) gauge fields $A_{\mu}$ in commutative spacetime, 
where $\eta$ is the conformal time $\eta = \int dt/a(t)$.  
In the case of power-law inflation models  
with the exponential inflaton potential in Eq.\ (\ref{eq:15}), 
it follows from (1) that the action for $A_i (k, \eta)$ in 
the Coulomb gauge, 
$A_0(t,\Vec{x}) = 0$ and ${\partial}_jA^j(t,\Vec{x}) =0$, 
can be expressed as follows.  
\begin{eqnarray} 
S^{(\mathrm{EM})} = \int
d{\eta} d^3k \frac{1}{2} b^2
\left[ {{A_i}^{\prime}}^{*}(k, \eta) {A_i}^{\prime}(k, \eta) 
       - k^2 {A_i}^{*}(k, \eta) A_i (k, \eta)
\right],  
\label{eq:35} 
\end{eqnarray}  
where a prime denotes derivative with respect to the conformal 
time $\eta$, and 
\begin{eqnarray}
&&b^2 = f(a) = \bar{f} a^{\beta-1} = \bar{f} a^{2q}, 
\label{eq:36} \\[3mm]
&&q \equiv 
\left( \frac{\tilde\beta-1}{2} \right) \left( \frac{p-1}{p} \right).  
\label{eq:37} 
\end{eqnarray}  
In deriving the expression of $q$ in (\ref{eq:37}), 
we have used Eq.\ (\ref{eq:21}).  
By introducing spacetime noncommutativity into the action (\ref{eq:35}), 
the modified action can be expressed as follows.  
\begin{eqnarray}
S_{\mathrm{SSUR}}^{(\mathrm{EM})} = \int_{k < k_{\mathrm{max}}}
d{\tilde\eta} d^3k \frac{1}{2} {z_k}^{2}(\tilde\eta) 
\left[ {{A_i}^{\prime}}^{*}(k, \tilde\eta) {A_i}^{\prime}(k, \tilde\eta) 
       - k^2 {A_i}^{*}(k, \tilde\eta) A_i (k, \tilde\eta)
\right], 
\label{eq:38} 
\end{eqnarray}
where 
\begin{eqnarray}
&&{z_k}^2(\tilde\eta) = b^2 (\beta^-_k \beta^+_k)^{1/2}.  
\label{eq:39}
\end{eqnarray}  
This action is similar to that for scalar and tensor 
fluctuation (\ref{eq:30}).  
As in the case of the scalar and tensor fluctuations, 
introducing spacetime noncommutativity corresponds to including 
the factor $(\beta^-_k \beta^+_k)^{1/2}$ in the action (\ref{eq:38}) 
as ${z_k}^2$.  
From (\ref{eq:38}) the equation of motion for $A_i (k, \tilde\eta)$ 
reads 
\begin{eqnarray}
{A_i}^{\prime\prime} (k, \tilde\eta) 
+ 2 \frac{ {z_k}^{\prime} }{z_k} {A_i}^{\prime} (k, \tilde\eta) 
        + k^2 {A_i}(k, \tilde\eta) = 0.  
\label{eq:40}
\end{eqnarray}
The friction term can be eliminated via a change of variables 
\begin{eqnarray}
&&u_i (k, \tilde\eta) = {z_k}(\tilde\eta) {A_i}(k, \tilde\eta), 
\label{eq:41}  
\end{eqnarray}
yielding the equation of motion 
\begin{eqnarray}
{u_i}^{\prime\prime} (k, \tilde\eta) 
+ \left( k^2 - \frac{{z_k}^{\prime\prime}}{z_k}  \right) 
{u_i} (k, \tilde\eta) = 0.  
\label{eq:42} 
\end{eqnarray} 
This equation is of the same type as that for scalar and tensor 
perturbations.  
Hence we can consider the power spectrum of $A_{\mu}$ in the same way as 
that of scalar and tensor perturbations.   

Here we introduce the following new parameter $\mu$ describing the 
noncommutativity of spacetime 
\begin{eqnarray}
\mu \equiv \frac{k^2 L_s^4}{\tau^2}, 
\label{eq:43} 
\end{eqnarray} 
and consider the case $\mu \ll 1$, which is the equivalent limit 
to the above UV one $k \gg k_{\rm crit}$, that is, we consider 
the quasi UV modes.  For these modes we can derive the approximate 
analytic solutions of Eq.\ (\ref{eq:42}) as argued below.  
Substituting  Eqs.\ (\ref{eq:29}) and (\ref{eq:36}) into 
Eq.\ (\ref{eq:39}) and using Eq.\ (\ref{eq:33}), we find 
\begin{eqnarray}
&& z_k \approx {\bar{f}}^{1/2} 
\bar{\alpha}^q {\tau}^{y_1} \left( 1 + \frac{y_2}{2}\mu \right), 
\label{eq:44} \\[3mm] 
&& y_1 \equiv \frac{pq}{p+1},  \hspace{5mm} 
y_2 \equiv \frac{y_1}{q} = \frac{p}{p+1},
\label{eq:45}
\end{eqnarray} 
where we have only recorded the terms up to the first order of $\mu$.  
Calculating ${z_k}^{\prime\prime}$ 
with the relation between $d\tilde\eta$ and $d\tau$ (\ref{eq:31}) 
and using Eqs.\ (\ref{eq:29}) and (\ref{eq:33}), 
we find 
\begin{eqnarray}
\frac{ {z_k}^{\prime\prime} }{z_k} \approx 
{\bar{\alpha}}^{4} y_1 (y_1 + 2y_2 -1) \tau^{4y_2 - 2}
\left[ 1 - \frac{y_2(2{y_1}^2 + 4y_1y_2 - 2y_1 + 2y_2 - 3)}
{y_1 (y_1 + 2y_2 -1)} \mu 
\right], 
\label{eq:46}  
\end{eqnarray} 
where we have taken the terms up to the first order of $\mu$.  
Furthermore, integrating the relation between $d\tilde\eta$ and $d\tau$ 
(\ref{eq:31}) and using Eqs.\ (\ref{eq:28}) and (\ref{eq:33}), we find 
\begin{eqnarray}
{\tau}^{4y_2 - 2} \approx 
\frac{1}{ {\bar{\alpha}}^{4} (1-2y_2)^2 } \frac{1}{{\tilde\eta}^2}, 
\label{eq:47}  
\end{eqnarray} 
where we have only recorded the terms up to the zeroth order of $\mu$.  
Finally, substituting Eq.\ (\ref{eq:47}) into Eq.\ (\ref{eq:46}) and 
taking the terms up to the zeroth order of $\mu$, we find 
\begin{eqnarray}
\frac{ {z_k}^{\prime\prime} }{z_k} \approx 
\frac{y_1(y_1 + 2y_2 -1)}{(1-2y_2)^2} \frac{1}{{\tilde\eta}^2}.  
\label{eq:48}  
\end{eqnarray} 
From Eqs.\ (\ref{eq:7}) and (\ref{eq:41}) 
the normalization condition for ${u_i} (k, \tilde\eta)$ reads
\begin{eqnarray} 
{u_i} (k, \tilde\eta) {{u_j}^{\prime}}^{*} (k, \tilde\eta) - 
{u_j}^{\prime} (k, \tilde\eta) {u_i}^{*} (k, \tilde\eta) 
= i \left( {\delta}_{ij} - \frac{k_i k_j}{k^2 } \right).  
\label{eq:49} 
\end{eqnarray} 
It follows from Eq.\ (\ref{eq:48}) that Eq.\ (\ref{eq:42}) is 
approximately rewritten to the following form.  
\begin{eqnarray}
{u_i}^{\prime\prime} (k, \tilde\eta) 
+ \left( k^2 - \frac{\nu^2-1/4}{ {\tilde\eta}^2 }  \right) 
{u_i} (k, \tilde\eta) = 0, 
\label{eq:50} 
\end{eqnarray} 
where
\begin{eqnarray}
\nu \approx \pm 
\frac{2(y_1 + y_2)-1}{2(1-2y_2)} 
= \mp \frac{1}{2} \tilde\beta.  
\label{eq:51} 
\end{eqnarray}
Here the second equality follows from Eqs.\ 
(\ref{eq:37}) and (\ref{eq:45}).  
The solution of Eq.\ (\ref{eq:50}) is given by 
\begin{eqnarray}
{u_i} (k, \tilde\eta) = 
C_{i+} (-\tilde\eta)^{1/2} H_{\nu}^{(1)} (-k\tilde\eta) + 
C_{i-} (-\tilde\eta)^{1/2} H_{\nu}^{(2)} (-k\tilde\eta),
\label{eq:52} 
\end{eqnarray} 
where $H_{\nu}^{(n)}$ is an $\nu$-th order Hankel function of type $n$ 
($n = 1,2$), 
and $C_{i+}$ and $C_{i-}$ are constants which satisfy 
\begin{eqnarray}
|C_{i+}|^2 - |C_{i-}|^2 = \frac{\pi}{4}.
\label{eq:53} 
\end{eqnarray} 
We shall choose
\begin{eqnarray}
C_{i+} = \frac{ \sqrt{\pi}}{2} e^{i(2\nu+1)\pi/4},  \hspace{5mm} 
C_{i-} = 0, 
\label{eq:54}
\end{eqnarray} 
so that the vacuum reduces to the one in Minkowski spacetime at the 
short-wavelength limit.  
Hence the power spectrum of the U(1) gauge fields $A_{\mu}$ in the vacuum state $|0 \rangle$ of the fields is 
\begin{eqnarray}
&&{\mathcal P}_{A_{\mu}}(k,\tilde\eta) = 2 \frac{k^3}{2\pi^2} 
\frac{ \langle 0|{u_i}^{*}(k,\tilde\eta){u_i}(k,\tilde\eta) |0 \rangle  }
     {{z_k}^2 (\tilde\eta)},  
\label{eq:55} 
\end{eqnarray}
where the factor 2 represents the degree of freedom of the transverse 
component of the U(1) gauge fields $A_{\mu}$.  

First, we consider the UV modes $k \gg k_{\rm crit}$, 
corresponding to the length scales smaller than $2.7 \times 10^2$Mpc today, 
which are generated within the Hubble radius and 
evolve in the same way as in the case of commutative spacetime.  
For these modes, 
$\beta^{\pm}_k (\tau) \approx a^{\pm 2}(\tau)$ and then 
$\tilde\eta \approx \eta$.  
Hence from Eq.\  (\ref{eq:39}) we find ${z_k}^{2} \approx b^2$.  
Expanding Eq.\  (\ref{eq:52}) in the long-wavelength limit and 
substituting the resultant expression and the relation 
${z_k}^{2} \approx b^2$ into Eq.\ (\ref{eq:55}), we obtain 
the expression for the power spectrum of the U(1) gauge fields 
in the inflationary stage as 
\begin{eqnarray}
{\mathcal P}_{A_{\mu}}(k,\eta) 
= 
\frac{2^{|\tilde{\beta}|-2}}{{\pi}^3} 
{\Gamma}^2 \left( \frac{|\tilde{\beta}|}{2} \right) 
\frac{1}{\bar{f} a^{\frac{(\tilde\beta-1)(p-1)}{p}} (\eta) }
\left[ \frac{p}{(p-1)a(\eta)H(\eta)}
\right]^{1-|\tilde{\beta}|} k^{3-|\tilde{\beta}|}, 
\label{eq:56} 
\end{eqnarray}
where we have taken $\nu = \tilde{\beta}/2$ and 
used the relation 
$
-\eta = p/[(p-1)aH]
$.  
Since the power spectrum of magnetic fields is 
${\mathcal P}_{B}(k,\eta) \propto k^2 {\mathcal P}_{A_{\mu}}(k,\eta)$, 
from Eq.\ (\ref{eq:56}) we find 
\begin{eqnarray}
{\mathcal P}_{B}(k,\eta) \propto k^{n_{\mathrm{UV}}}, \hspace{5mm} 
n_{\mathrm{UV}} = 5-|\tilde\beta|.  
\label{eq:57} 
\end{eqnarray}
As a result, the dependence of the power spectrum of magnetic fields 
on the comoving wavenumber $k$ is the same as that in the case of commutative 
spacetime in Eq.\ (\ref{eq:22}),  
and hence the SSUR has no significant effect on the UV modes, 
as expected.  

Furthermore, after inflation 
the relation between the energy density of the 
magnetic fields in the position space 
${\rho}_{B}(L,t)$ 
and the power spectrum of the U(1) gauge fields 
${\mathcal P}_{A_{\mu}}(k,\eta)$ is given by 
\begin{eqnarray}
{\rho}_{B}(L,t) 
= \frac{1}{2} \left( \frac{1}{a} \right)^2
\left( \frac{k}{a} \right)^2 {\mathcal P}_{A_{\mu}}
(k,{\eta}_\mathrm{R}) f(a), 
\label{eq:58}
\end{eqnarray}
where ${\mathcal P}_{A_{\mu}}(k,{\eta}_\mathrm{R})$ is the 
power spectrum of the U(1) gauge fields at the end of inflation.  
Here we have taken into account the fact that 
since after inflation the conductivity of the Universe becomes 
a value much larger than the Hubble parameter at that time, 
the power spectrum of the U(1) gauge fields freezes at the end of 
inflation.  
It follows from Eqs.\ (\ref{eq:56}) and (\ref{eq:58}) 
that 
the energy density of the magnetic fields for the UV modes 
${\tilde{{\rho}_{B}}}^{(\mathrm{UV})}$
at the present time is given by 
\begin{eqnarray}
{\tilde{{\rho}_{B}}}^{(\mathrm{UV})}(L,t_0) 
   &=&
  \frac{2^{|\tilde{\beta}|-3}}{{\pi}^3} 
  {\Gamma}^2 \left( \frac{|\tilde{\beta}|}{2} \right)
  \left( \frac{p-1}{p} \right)^{ |\tilde{\beta}| - 1}
   {H_{\mathrm{R}}}^4 
   \left( \frac{a_{\mathrm{R}}}{a_0}  \right)^4 
  { \left( \frac{k}{a_{\mathrm{R}} H_{\mathrm{R}}} \right) }^
                                                {- |\tilde{\beta}| + 5} 
\nonumber  \\[2.5mm]
&& \hspace{55mm} \times 
  \exp \left( \hspace{0.5mm}  - \tilde{\lambda} \kappa {\Phi}_\mathrm{R}
          X  \hspace{0.5mm} \right) 
  \left(  \Delta \tilde{S}   \right)^{-4/3}, 
\label{eq:59} 
\end{eqnarray}
where we have taken into account the evolution of the dilaton after reheating 
and the entropy production by that decay in the same way as done 
in the case of commutative spacetime.  
This expression is equivalent to that for 
the energy density of large-scale 
magnetic fields at the present time in commutative spacetime 
$\tilde{{\rho}_{B}}(L,t_0)$ in (\ref{eq:22}).  
Consequently, 
the above expression of the strength of the generated magnetic fields 
on scales smaller than $2.7 \times 10^2$Mpc today 
in noncommutative spacetime is 
equivalent to that in the commutative one.  
We can therefore yield sufficiently strong magnetic fields $B \sim 10^{-10}$G 
on 1Mpc scale for reasonable values of $H_{\mathrm{R}}$ and $\Delta \tilde{S}$ 
as before, provided that the spectrum is nearly scale invariant, 
$\tilde\beta \approx 5$.  

As we argued in the previous subsection, the important consequence of the 
SSUR is that the power-law index, $p$, of the power-law inflation could 
be much smaller than in the case of commutative geometry without conflicting 
with the nearly scale-invariant spectrum of density fluctuations observed by 
WMAP. Then from Eq.\ (\ref{eq:21}) and, say, $p \approx 5$, which is 
in the allowed range now, we find 
\begin{eqnarray}
\tilde\beta = 1 + \frac{X}{2}.  
\label{eq:60} 
\end{eqnarray}  
Hence we can generate magnetic fields as strong as $B \sim 10^{-10}$G on 1Mpc 
scale even if the two coupling constants of the dilaton, $\lambda$ and 
$\tilde\lambda$, are of the same order of magnitude, 
$X = \lambda / \tilde\lambda \approx 8$.  
For example, we find $B(1\mathrm{Mpc},t_0) = 1.0 \times 10^{-10} \mathrm{G}$ 
for $X=8.1$, $H_\mathrm{R}=10^{8}$GeV, and 
$\Delta \tilde{S} = 7.0 \times 10^{6}$.  
Moreover, 
we find similarly that 
a sufficient magnitude of magnetic fields on 1Mpc scale at the present time 
for the galactic dynamo scenario, $B \sim 10^{-22}$G, could be generated 
even in the case $X = \lambda / \tilde\lambda \approx 6$.  

For completeness 
we also consider the IR modes $k \ll k_{\rm crit}$, which are generated 
outside the Hubble radius.  
It follows from Eq.\ (\ref{eq:27}) that 
the time when these modes are generated 
is given by \cite{Brandenberger} 
\begin{eqnarray}
{\tau}_k = 
\left[ 
k^2 {L_s}^4 +  \frac{ \left( k L_s \right)^{2(p+1)/p} }{(p+1)^2 \bar{a}^{2/p}}
\right]^{1/2}.
\label{eq:61} 
\end{eqnarray}  
For the IR modes $k \ll k_{\rm crit}$, the first term dominates over 
the second one and hence ${\tau}_k \approx k {L_s}^2$. 
In this case, from Eqs.\ (\ref{eq:29}) and (\ref{eq:33}) 
we find 
\begin{eqnarray}
\beta^{\pm}_k (\tau) \approx \frac{1}{2} a^{\pm 2}(\tau \pm {L_s}^2 k).  
\label{eq:62} 
\end{eqnarray} 
Substituting Eqs.\ (\ref{eq:36}) and (\ref{eq:62}) 
into Eq.\ (\ref{eq:39}) and taking the first leading term in the 
limit $k \ll k_{\rm crit}$, 
we find 
\begin{eqnarray}
{z_k}^2({\tau}_k) \approx \frac{1}{2} \bar{f} 
a^{2q} (k {L_s}^2) 
a^{-1} \left( 
\frac{k(k L_s)^{2/p}}{2(p+1)^2 \bar{a}^{2/p}}
\right) 
a (2 k {L_s}^2).  
\label{eq:63} 
\end{eqnarray} 
Moreover, it follows from the relation between $d\tilde\eta$ and $d\tau$, 
(\ref{eq:31}), 
that 
the relation between $\tilde{\eta}$ and $\tau$ at 
the time of the generation of 
the IR modes is approximately given by 
\begin{eqnarray}
\tilde{\eta}_k 
&\approx& -a^{-1} \left( 
\frac{k(k L_s)^{2/p}}{2(p+1)^2 \bar{a}^{2/p}}
\right) a^{-1} (2 k {L_s}^2) {\tau}_k  \nonumber \\[2.5mm] 
&\approx& - \frac{1}{k},  
\label{eq:64} 
\end{eqnarray}
where we have used the equality ${\tau}_k \approx k {L_s}^2$.  
Expanding Eq.\  (\ref{eq:52}) in the long-wavelength limit and 
substituting the resultant expression at the time of the generation 
${\tilde\eta}_k \approx -1/k$ and Eq.\ (\ref{eq:63}) into Eq.\ (\ref{eq:55}), 
we obtain 
\begin{eqnarray}
{\mathcal P}_{A_{\mu}}(k,{\tilde\eta}_k) = C_{\mathrm{IR}} 
k^{ \frac{1}{p+1} \left[ 2(p+2)- (\tilde\beta-1)(p-1)\right] },  
\label{eq:65} 
\end{eqnarray}
where
\begin{eqnarray}
C_{\mathrm{IR}} = 
\frac{2^{ |\tilde\beta| - \frac{3p+1}{p+1}}}{\pi^3 \bar{f}} 
{\Gamma}^2 \left( \frac{|\tilde{\beta}|}{2} \right) 
\left[ (p+1)^p \bar{a} 
\right]^{ -\frac{2}{p+1} \left[ \frac{(\tilde\beta-1)(p-1)}{2p} + 1\right] } 
{L_s}^{ -\frac{2\tilde\beta (p-1)}{p+1} }.  
\label{eq:66} 
\end{eqnarray}
Furthermore, from Eq.\ (\ref{eq:65}) and 
${\mathcal P}_{B}(k,\tilde\eta) \propto k^2 {\mathcal P}_{A_{\mu}}
(k,\tilde\eta)$ 
we find 
\begin{eqnarray}
{\mathcal P}_{B}(k,{\tilde\eta}_k) \propto 
k^{ n_{\mathrm{IR}} }, \hspace{5mm} 
n_{\mathrm{IR}} = \frac{1}{p+1} \left[ 2(2p+3)- (\tilde\beta-1)(p-1)\right].  
\label{eq:67}
\end{eqnarray}
It follows from Eqs.\ (\ref{eq:57}) and (\ref{eq:67}) that 
$n_{\mathrm{IR}} - n_{\mathrm{UV}} = 2 \tilde\beta /(p+1) > 0$, 
where we have taken $\tilde\beta > 0$.  
As a result, only the power for the long-wavelength modes 
larger than the crossover scale $\sim 2.7 \times 10^2$Mpc 
tends to be suppressed.

\section{Conclusion}
In the present paper we have discussed a possible solution to the huge 
hierarchy between $\lambda$ and $\tilde\lambda$, which is required 
in our previous work \cite{Bamba} 
in order that the spectrum of generated magnetic field should not be too 
blue but close to the scale-invariant or the red one, so that 
the amplitude of the generated magnetic field could be sufficiently large, 
by taking account of the effects of the SSUR on the primordial power 
spectrum of metric perturbations in the early Universe.  
As a result we have found that 
in power-law inflation models, owing to the consequences of the SSUR on 
metric perturbations, 
the resultant magnetic fields could have a nearly scale-invariant spectrum 
even in the case 
$\lambda$ and $\tilde\lambda$ are of the same order of magnitude, 
$X = \lambda/\tilde{\lambda} \approx 8$, 
so that the amplitude of the generated magnetic field could be 
as large as $10^{-10}$G on 1Mpc scale at the present time, which is strong 
enough to account for the observed fields in galaxies and clusters of galaxies 
through only adiabatic compression 
without requiring any dynamo amplification.  
Since the strength of the magnetic fields on megaparsec scales is expressed 
by the same formula as in the case of commutative geometry, 
this result is entirely due to the fact that 
in the presence of the SSUR the power index of power-law inflation 
could be much smaller than the case of the commutative geometry in order to 
reproduce the nearly scale-invariant spectrum of density fluctuation observed 
by WMAP.

\section*{Acknowledgements}
This work was partially supported by 
the Monbu-Kagaku Sho 21st century COE Program 
``Towards a New Basic Science; Depth and Synthesis" 
and J.Y.\ was also supported in part by the JSPS Grant-in-Aid for 
Scientific Research Nos.13640285 and 16340076.


\end{document}